\documentstyle[prl,twocolumn,aps,pstricks]{revtex} 
\begin{document} 
\draft 
\title{Direct Probing of Quantum Phase Space  by Photon Counting} 
\author{Konrad Banaszek and Krzysztof W\'odkiewicz\cite{unm}} 
\address{Instytut Fizyki Teoretycznej, Uniwersytet Warszawski, 
00--681 Warszawa, Ho\.za 69, Poland} 
\date{February 29, 1996} 
\maketitle 
\begin{abstract} 
We propose a very simple experimental setup to measure, via 
photon counting, the overlap of the Wigner functions characterizing 
two single mode light beams. We show that this scheme can be 
applied to determine directly the phase space quasiprobability distribution 
of the single mode field and in a certain limit the Wigner 
function can be measured without use of tomographic 
reconstruction algorithms. The deleterious effects of non--unit 
photodetector efficiency are analyzed. 
\end{abstract} 
The concept of phase space, fundamental in classical 
mechanics, remains useful when passing to quantum theory. It is 
well known that  quantum states can be fully characterized by 
the Wigner function defined in the phase space and that the quantum 
expectation values can be represented as statistical averages of the phase 
space variables \cite{PhaseSpaceReview}.  In quantum optics this formalism 
provides a convenient framework for 
discussion of many topics \cite{VogelWelsch}.

An exciting problem explored 
recently is the experimental determination of the Wigner function of 
a single light mode. It was first shown theoretically by Vogel 
and Risken \cite{VogeRiskPRA89} that quadrature amplitude 
distributions measured in homodyne detection provide enough data 
to perform a complete reconstruction of the Wigner function. This 
method, called quantum tomography, was successfully realized  
in a series of experiments \cite{tomography}.  The quantity recorded in 
the quantum tomography  experiments was the statistics of the count
difference of photodetectors facing the signal field superimposed on
a strong  coherent field. The Wigner function was reconstructed from 
these data using  numerical algorithms of the inverse Radon transform.  There
are also  other known methods for the
 complete experimental characterization of
the single mode state: heterodyne \cite{ShapWagnIEE84} and double homodyne 
\cite{PhaseInPhaseSpace,WalkJMO87,DblHomExp} detection, where the  
so--called $Q$ function of the signal field, which is a smoothed 
Wigner function, is measured. 

In this Letter we propose an alternative scheme for the phase space 
measurement. Given two single mode light beams we present an 
extremely simple experimental setup to measure the overlap of 
the Wigner functions characterizing these field. Moreover we show that the 
Wigner functions can be relatively rescaled by an arbitrary 
positive factor. The measurement is performed by optical means 
and only a trivial arithmetic operation has to be done on the 
data recorded from the photodetector. 

Before we discuss the proposed measurement, we review briefly, using a
Heisenberg-picture-type of approach, the properties of the
phase space distributions used in this Letter. 
The Wigner function or the $Q$ function are examples of more general 
$s$-parameterized quasiprobability distributions $W(\alpha;s)$ in 
the complex $\alpha$ phase space \cite{CahilGlauber}, with $s$=$0$
corresponding to the Wigner function $W(\alpha)=W(\alpha;0)$. These
distributions 
can be written as  expectation values of the following normally 
ordered operator: 
\begin{equation} 
\label{U(a,s)} 
{\hat U}( \alpha; s)= 
 \frac{2}{ \pi (1-s)}\ : \exp \left( - 
\frac{2}{1-s}(\alpha^{*}- \hat{a}^\dagger) (\alpha- \hat{a} ) 
\right) : 
\end{equation} 
where $\hat{a}^\dagger$ and $\hat{a}$ are the single mode photon 
creation and annihilation  operators. In the proposed measurement scheme
we utilize the simple fact that the single mode Wigner function at the origin
of the complex phase space $W(0)$ can be directly computed from the
distribution of counts measured by a photodetector facing this mode. 
Indeed for $\alpha=0$ and $s=0$  the
operator (\ref{U(a,s)}) can be written in the 
following equivalent forms: 
\begin{eqnarray} 
\label{U(0,0)} 
{\hat U}(0;0) &  = & \frac{2}{\pi}  : \exp \left( - 2 
\hat{a}^\dagger \hat{a} \right) : \nonumber \\ 
& = & \frac{2}{\pi} \sum_{n=0}^{\infty} (-1)^n  : 
e^{- \hat{a}^\dagger \hat{a}} \frac{(\hat{a}^\dagger 
\hat{a})^n}{n!} :\nonumber \\ 
& = & \frac{2}{\pi}\sum_{n=0}^{\infty} (-1)^n | n \rangle 
\langle n|  
\end{eqnarray} 
where we have used the normally ordered operator representation 
of the $n$ photon number projection operator. The quantum 
expectation value of this operator gives: 
\begin{equation} 
\label{W(0)=sum} 
W(0) = \langle {\hat U}(0;0) \rangle = \frac{2}{\pi} 
\sum_{n=0}^{\infty} 
(-1)^n p_{n} 
\end{equation} 
where the value $p_{n}$ appearing 
in this expansion is just the probability of counting $n$ 
photons  by an ideal photodetector. 
Thus the photoncount statistics allows to calculate the Wigner 
function at the origin of the phase space. It has been pointed 
out by Royer \cite{RoyePRL85} that the ability to measure
$W(0)$ allows us to scan the complete Wigner function by shifting 
the system or equivalently the frame of reference in the phase 
space. Although in principle this shifting of the reference frame might be 
realized in quantum optics, its experimental realization would encounter 
difficulties. Therefore we will utilize the results of the 
measurement of $W(0)$ in an alternative way, providing a much more 
feasible experimental scheme. The proposed scheme is very general and we
believe that it may find many various applications and generalizations. In
the present Letter we will discuss in detail one specific example motivated
by the quantum tomography experiments. 

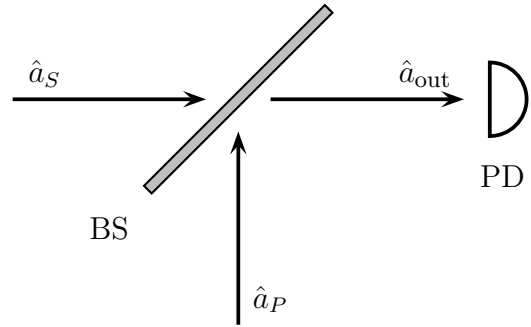
\begin{figure} 
\begin{center}
\psset{unit=.3375in}
\begin{pspicture}(10,8)
\rput{45}(4.5,4){%
\psframe[fillstyle=solid,fillcolor=lightgray](-2,-.1)(2,.1)%
} 
\psset{linewidth=.5mm}
\psline{->}(1,4)(4,4)
\psline{->}(5,4)(8,4)
\psline{->}(4.5,.5)(4.5,3.5)
\pswedge(8.4,4){.6}{270}{90}
\rput(2.5,2){\large BS}
\rput(8.6,2.8){\large PD}
\rput(1.5,4.4){\large $\hat{a}_S$}
\rput(7.4,4.4){\large $\hat{a}_{\mbox{\small\rm out}}$}
\rput(5.0,0.9){\large $\hat{a}_P$}
\end{pspicture}
\end{center}
\caption{Experimental setup discussed in the Letter. BS denotes the 
beamsplitter, PD is the photodetector, and the annihilation 
operators of the modes are indicated.\label{Fig1}} 
\end{figure}

The proposed setup is presented in Fig.\ \ref{Fig1}. 
We will take the detected field to be a superposition of two 
single mode fields, which we will call the signal and the probe, 
and denote their annihilation operators by $\hat{a}_S$ and 
$\hat{a}_P$ respectively. Such a combination can be easily 
realized by means of a beamsplitter.   As it is well known 
\cite{VogelWelsch}, the action of the beamsplitter is 
described by an $\text{SU}(2)$ transformation between the 
annihilation operators of the incoming and outgoing modes. 
Since the phase shifts appearing 
in this transformation can be eliminated by an appropriate 
redefinition of the modes, the annihilation operator of the 
outgoing mode falling onto the detector face 
$\hat{a}_{\text{out}}$ can be assumed to be a combination 
\begin{equation} 
\hat{a}_{\text{out}} = \sqrt{T} \hat{a}_S - \sqrt{1-T} \hat{a}_P 
\end{equation} 
where $T$ is the beamsplitter power transmissivity
The Wigner function of the outgoing mode at the phase space 
origin is given in terms of the incoming modes by the 
expectation value of 
\begin{eqnarray} 
{\hat U}_{\text{out}}(0;0) & = & \frac{2}{\pi} \, : 
\exp \left( - 2 
\hat{a}^\dagger_{\text{out}} \hat{a}_{\text{out}} \right) : 
\nonumber \\ 
& = & \frac{2}{\pi} \, : \exp \left( - 2 T 
(\hat{a}^{\dagger}_{S}- \sqrt{(1-T)/T}\hat{a}^{\dagger}_{P}) 
\right. 
\nonumber \\ 
\label{U=SP} 
&   & \times \left. ( \hat{a}_{S} 
- \sqrt{(1-T)/T} \hat{a}_{P} ) \right) :  \;  .
\end{eqnarray} 
This simple relation provides an interesting link between the 
detected quantity and the $S$ mode. Let us for example consider 
a simple case when the probe field is a coherent state 
$\hat{a}_P | \alpha \rangle = \alpha | \alpha \rangle$ 
uncorrelated with the signal mode. Performing the quantum 
average over the $P$ mode in Eq.\ (\ref{U=SP}) is straightforward 
due to the normal ordering of the operators. Taking the expectation 
value over the signal and using the definition (\ref{U(a,s)}) 
we obtain that  the Wigner function (\ref{W(0)=sum}) for the outgoing mode 
is proportional to an $s=1-1/T$ ordered  quasidistribution function of the
$S$ mode: 
\begin{equation} 
\label{Wout(0)=W_S} 
W_{\text{out}}(0)= \frac{1}{T} W_S \left( \sqrt{\frac{1-T}{T}} 
\alpha ; - \frac{1-T}{T} \right). 
\end{equation} 
Thus our setup delivers directly the value of the signal 
quasidistribution function at the phase space point dependent on the 
amplitude and the phase of the probe coherent state. Since both 
these parameters can be controlled experimentally without
difficulties, we may simply scan the phase space by changing the 
amplitude and the phase of the probe field and thus determine the 
complete quasidistribution function. Eq.\ (\ref{Wout(0)=W_S}) 
shows that its ordering depends on the beamsplitter 
transmissivity and for $T$ near one approaches zero, which means 
that the scanned quasidistribution is close to the Wigner 
function of the signal field. 

We shall now generalize Eq.\ (\ref{Wout(0)=W_S}) for arbitrary 
state of the probe mode by 
considering the disentanglement of the $S$ 
and $P$ modes. In order to achieve this  we use 
the following Gaussian integration of normally ordered operators 
of the $S$ and $P$ modes: 
\begin{eqnarray} 
\label{U=S+P} 
\lefteqn{\hat{U}_{\text{out}}(0;0)} & & \nonumber \\ 
& = & \frac{4}{\pi^2}  \int \text{d}^2 \beta \, 
: \exp \left( - 2 (\sqrt{T} \beta^{*} - \hat{a}^{\dagger}_{P} ) 
(\sqrt{T} \beta - \hat{a}_{P} ) \right): 
\nonumber \\ 
&  &  \times : \exp \left( - 2 
(\sqrt{1-T}\beta^{*} -\hat{a}^{\dagger}_{S} ) 
(\sqrt{1-T}\beta - \hat{a}_{S} ) \right) : \; . 
\nonumber \\ 
& & 
\end{eqnarray} 
Under the assumption that the $S$ and $P$ modes are uncorrelated 
this disentanglement yields the following expression for the 
quantity detected by our setup: 
\begin{eqnarray} 
\label{W(0)=int} 
W_{\text{out}}(0) & = & \int \text{d}^2 \beta \, W_S(\sqrt{1-T} 
\beta) W_P (\sqrt{T} \beta) \nonumber \\ 
& = & \frac{1}{1-T} \int \text{d}^2\beta \, W_S(\beta) W_P 
(\sqrt{T/(1-T)} \beta). 
\end{eqnarray} 
This formula establishes the connection between the photon statistics 
of the outgoing mode and the Wigner functions 
of the $S$ and $P$ modes. It reflects the fundamental
advantages of
our setup in the direct probing of the phase space of the light field.
In the case when the beamsplitter splits the light equally, 
we have $\sqrt{T/(1-T)} = 1$ and $W_{\text{out}}(0)$ is simply a
doubled overlap of the signal and probe Wigner functions. In the general 
case the phase space parameterization of the probe Wigner 
function is rescaled by the factor $\sqrt{T/(1-T)}$ which can 
take an arbitrary positive value depending on the beamsplitter 
transmissivity. This rescaling causes an effective decrease of the probe
width by the factor $\sqrt{(1-T)/T}$ in {\it 
all\/} the quadratures simultaneously.  This contrasts with the 
unbalanced double homodyne detection scheme 
\cite{WalkJMO87,LeonPRA93}, where the resolution of the phase 
space probing along one quadrature can be improved only at the 
cost of deteriorating the resolution along the perpendicular 
direction. For the coherent probe field $|\alpha\rangle$ the $P$ 
mode Wigner function is of the form 
\begin{equation} 
\label{WPcoh} 
W_P(\beta) = \frac{2}{\pi} \exp \left ( -2 | \beta - \alpha | ^2 
\right). 
\end{equation} 
An easy computation shows that in this case 
Eq.\ (\ref{W(0)=int}) indeed reduces to 
Eq.\ (\ref{Wout(0)=W_S}). The rescaling of the probe Wigner 
function in the convolution (\ref{W(0)=int}) has another 
consequence.  When $T$ tends to one, the factor multiplying the 
probe amplitude $\alpha$ becomes very small and to scan the 
interesting region of the signal phase space one has to use a 
probe field of large intensity.  

One of the advantages of balanced homodyne detection 
used in quantum tomography experiments is the cancellation of the excess
noise of the reference field. In our scheme
this noise deteriorates the resolution 
with which the signal phase space can be probed.
This can become important
in the limit $T\rightarrow 1$, where strong probe fields have
to be used. The influence of the excess noise can be simply estimated 
assuming a Gaussian thermal noise  described by the following  Wigner
function
\begin{equation}
W_P(\beta) = \frac{2}{\pi(2\bar{n} +1)} \exp
\left( - \frac{2}{2\bar{n}+1} | \beta - \alpha|^2 \right),
\end{equation}
where $\bar{n}$ is the mean  number of thermal
photons in the beam. 
An easy calculation shows that for such a noisy probe field
\begin{equation}
W_{\text{out}} (0) = \frac{1}{T} W_S \left( \sqrt{\frac{1-T}{T}}
\alpha ; - (2\bar{n}+1) \frac{1-T}{T} \right).
\end{equation}
In particular, when $\bar{n} \gg 1$ grows linearly with the probe
intensity $|\alpha|^2$, the excess noise imposes a restriction
on the highest ordering of the signal quasidistribution function
measured at a given point.

The proposed setup is an optical realization of a model scheme of 
quantum measurement \cite{WodkPRL84,BuzeKeitPRA95}, where in 
addition to the system a filter device---a ``quantum ruler''---is 
introduced and the measured phase space probability distribution 
is the  convolution of the system and filter Wigner functions. Our 
scheme is more general, since the Wigner function of the 
filter can be rescaled by an arbitrary factor. Consequently the 
rescaled probe Wigner function does not have to obey the 
Heisenberg uncertainty principle and may even 
approach the shape of a delta function, which leads to the direct 
measurement of the Wigner function. In contrast to quantum 
tomography no sophisticated computer processing of the 
experimental data is necessary. The quantity measured in the 
experiment is proportional to the quasiprobability distribution 
at the phase space point depending only on the amplitude and phase of the 
probe state.

In the remaining discussion we will introduce two generalizations. 
First we will make our considerations more realistic by taking 
into account the imperfectness of the photodetector. When the 
detector efficiency is $\eta$, the probability of counting $n$ 
photons is given by the expectation value of $:\exp( - 
\eta\hat{a}_{\text{out}}^\dagger 
\hat{a}_{\text{out}}) (\eta 
\hat{a}_{\text{out}}^\dagger \hat{a}_{\text{out}} )^n 
/ n! : $. The second extension is the substitution of the factor 
$(-1)^n$ in Eq.\ (\ref{W(0)=sum}) by $-(s+1)^n/(s-1)^{n+1}$, 
where $s$ is a real parameter. The origin and role of the 
parameters $\eta$ and $s$ is different: $\eta$ describes 
experimental limitations, while $s$ is an artificial number 
introduced in the numerical processing of the measured data. 
With these two parameters we obtain the following simple 
generalization of the formula (\ref{U=SP}), when expressed in 
terms of the $S$ and $P$ modes 
\begin{eqnarray} 
\lefteqn{{\hat U}_{\text{out}}^{(\eta)}(0,s)} 
& & 
\nonumber \\ 
&=& \frac{2}{\pi (1-s)} \sum_{n=0}^{\infty} \left( 
\frac{s+1}{s-1} \right)^{n} : 
e^{- \eta\hat{a}_{\text{out}}^\dagger \hat{a}_{\text{out}}} 
\frac{(\eta\hat{a}_{\text{out}}^\dagger 
\hat{a}_{\text{out}})^{n}}{n!} : 
\nonumber \\ 
& = & \frac{2}{\pi (1-s)} : 
\exp \left( - \frac{2\eta}{1-s} 
\hat{a}_{\text{out}}^\dagger \hat{a}_{\text{out}}\right) : 
\nonumber \\ 
& = & \frac{2}{\pi (1-s)} \  : \exp \left( - \frac{2\eta T}{1-s} 
(\hat{a}^{\dagger}_{S}-  \sqrt{(1-T)/T}\hat{a}^{\dagger}_{P} ) 
\right. 
\nonumber \\ 
\label{U(eta)} 
& & \;\;\;\;\;\; \times \left. \vphantom{\frac{2\eta T}{1-s}} 
( \hat{a}_{S} - \sqrt{(1-T)/T} \hat{a}_{P} ) \right) : \; .
\end{eqnarray} 
The third line of this equation suggests that the parameter $s$ 
can be used to compensate the imperfectness of the 
photodetector. Indeed if we selected $s =1 - \eta$, we would 
determine the expectation value of $: \exp ( - 2 
\hat{a}_{\text{out}}^\dagger \hat{a}_{\text{out}}):$ regardless 
of the detector efficiency. But in this case the factor 
multiplying the probability of counting $n$ photons is 
$(1-2/\eta)^n$ and its magnitude diverges to infinity with 
$n \rightarrow \infty$.  This is not important from a 
theoretical point of view, since we have shown that the series 
converges to the expectation value of a well behaved operator. 
However it becomes crucial in the processing of the measured 
probability distribution, which is influenced by experimental 
errors and statistical fluctuations, and hence need not tend to 
zero sufficiently quickly to assure the convergence of the 
complete series.  This reasoning might be opposed since the 
experimental sample of the photodetector counts is finite and 
thus the counts distribution is zero above certain photon 
number. Nevertheless the problem still exists, since the 
increasing factor in the generalization of the sum 
(\ref{W(0)=sum}) causes that an important contribution comes 
from the ``tail'' of the experimental counts distribution, which 
has usually has a very poor statistics, and consequently the 
final result has a huge statistical error 
\cite{RecovPhotStat}.  The simplest way to 
avoid all these problems is to assume the factors multiplying 
the counts statistics to be bounded, which is equivalent to the 
condition $s \leq 0 $. 

An easy computation shows that if the coherent state (\ref{WPcoh}) 
is employed as a probe, the expectation value of the generalized 
operator $\hat{U}_{\text{out}}^{(\eta)}(0,s)$ is again given by the 
quasidistribution function of the signal mode: 
\begin{equation} 
 \langle \hat{U}^{(\eta)}_{\text{out}} (0,s) \rangle 
= \frac{1}{\eta T} W_S \left( 
\sqrt{\frac{1-T}{T}} \alpha ; - \frac{1-s-\eta T}{\eta T} 
\right). 
\end{equation} 
Let us now analyze the ordering of this function. It has been pointed 
out by Leonhardt and Paul \cite{LeonPaulPRL94} that 
although from a theoretical point of view an arbitrarily ordered 
distribution contains the complete characterization of the quantum 
state, experimental errors 
make it difficult to compute higher ordered distributions from the
measured one.
Thus what is interesting is the highest ordering 
achievable in our scheme. 
Analysis of the role of the parameter $s$ is the simplest, 
since the greater its value, the higher is the ordering obtained. 
Because it is restricted in its range to nonpositive values, it 
should be consequently set to zero. Thus we are left with two 
parameters: $\eta$ and $T$. 
It is easy to check that for fixed $\eta$ 
the highest ordering is still achieved when $T\rightarrow 1$, 
but its limit value is now $-(1-\eta)/\eta$. Under the 
assumption that $\eta$ and $T$ are close to one, the ordering of 
the measured distribution is effectively equal to this limiting 
value if the difference $1-T$ is much smaller than $1-\eta$. For 
currently used photodetectors, this condition can be realized 
experimentally.  Thus the highest ordering achievable in our 
scheme is effectively determined by the photodetector efficiency 
and is equal to $-(1-\eta)/\eta$. It is noteworthy that this is 
exactly equal to the ordering of the distribution reconstructed 
tomographically from data measured in the homodyne detection 
with imperfect detectors \cite{LeonPaulPRA93}.

The measurement of the quasiprobability distribution does not exhaust 
possible applications of the proposed setup. Since the probe field may 
be in arbitrary state, the variety of information on the quantum state 
which can be retrieved using this scheme is potentially very large. 
Another interesting extension of the presented work is its 
generalization to the multimode case. 

We wish to acknowledge useful discussions with J. Mostowski and Cz.
Radzewicz. We thank  P. L. Knight for numerous comments 
about the final version of the manuscript. 
This work has been partially supported by the
Polish KBN grant.

{\it Note added.} After this Letter was submitted, 
Ref.\ \cite{Wallentowitz} was brought to the authors' attention. It discusses
an analogous measurement scheme with a coherent state used as a probe.

\end{document}